\documentstyle[floats,aps]{revtex}

\begin{document}

\newcommand{\de}{\delta}\newcommand{\ga}{\gamma}
\newcommand{\e}{\epsilon} \newcommand{\th}{\theta}\newcommand{\ot}{\otimes}
\newcommand{\ba}{\begin{array}} \newcommand{\ea}{\end{array}}
\newcommand{\beq}{\begin{equation}}\newcommand{\eeq}{\end{equation}}
\newcommand{\tmod}{{\cal T}}\newcommand{\amod}{{\cal A}}
\newcommand{\bemod}{{\cal B}}\newcommand{\cmod}{{\cal C}}
\newcommand{\dmod}{{\cal D}}\newcommand{\hmod}{{\cal H}}
\newcommand{\s}{\scriptstyle}\newcommand{\tr}{{\rm tr}}
\newcommand{\einsop}{{\bf 1}}
\title{Bethe ansatz solution of the  closed anisotropic
supersymmetric $U$ model with quantum supersymmetry.}
\author{Katrina Hibberd$^1$\thanks{keh@cbpf.br},
Itzhak Roditi$^2$\thanks{roditi@cbpf.br},
Jon Links$^3$\thanks{jrl@maths.uq.edu.au} and
Angela Foerster$^4$\thanks{angela@if.ufrgs.br}}
\date{November, 1999.}
\maketitle
\begin{center}
${}^{1,2}$Centro Brasileiro de Pesquisas Fisicas, \\
Rua Dr. Xavier Sigaud 150, 22290-180, Rio de Janeiro - RJ, Brazil.\\
\vspace{.5cm}
${}^3$Department of Mathematics, \\
The University of Queensland, 4072, Australia. \\
\vspace{.5cm}
${}^4$Instituto de Fisica da UFRGS,\\
Av. Bento Goncalves 9500, Porto Alegre, RS - Brazil.
\end{center}

\begin{abstract}
The nested algebraic Bethe ansatz is presented for the anisotropic
supersymmetric $U$ model maintaining quantum supersymmetry.
The Bethe ansatz equations of the model are obtained on
a one-dimensional closed lattice and an expression for the
energy is given.
\end{abstract}



\section{Introduction}

The study of strongly correlated electrons continues to receive a lot
of interest due to applications in condensed matter physics.
Some of the well known models are the Hubbard and t-J models 
\cite{korbook} and generalizations such as the 
 EKS model \cite{EKS}.  

Another example of a model describing strongly correlated electrons is the
the supersymmetric $U$ model.  This model
was first introduced in \cite{ww} and was shown to be integrable via
the Quantum Inverse Scattering Method (QISM) \cite{kul} by demonstrating
that the model could be obtained from an $R$-matrix which is invariant
with respect to the Lie superalgebra $gl(2|1)$. The Bethe-ansatz
equations for the model were obtained in \cite{periodic,meaba,marcio,frahm}.
Subsequently, an anisotropic generalization was presented in \cite{bar}
which was also shown to be integrable through use of an $R$-matrix derived
from a
representation of the quantum superalgebra $U_q(gl(2|1))$ \cite{me}.

The anisotropic supersymmetric $U$ model describes a system of
correlated electrons and generalizes the Hubbard model in the sense that
as well as the presence of the Hubbard on site (Coulomb) interaction
there are additional correlated hopping and pair hopping terms.
The model acts on the unrestricted $4^k$-dimensional electronic Hilbert
space $\otimes ^k_{n=1} C^4$ where $k$ is the lattice length.
This means that double occupancy of sites is allowed and
distinguishes this model from the anisotropic $t-J$ model \cite{angobc}
which
shares the same supersymmetry algebra $U_q(gl(2|1))$.
The model contains one free  parameter $U$, (the Hubbard
interaction parameter) which arises from the one-parameter
family of inequivalent typical four-dimensional irreducible
representations of the $U_q[gl(2|1)]$, and another which arises from the
deformation parameter $q$.

Bethe ansatz solutions for the anisotropic model with periodic boundary
conditions have been studied \cite{bar,mass,meqaba}, however for this
case there is no quantum superalgebra symmetry.  In
\cite{martin,GPPR,karow,kz,Angi}
some quantum algebra invariant integrable closed chains derived from
an $R$-matrix associated with the Hecke algebra were introduced
and investigated.  It was subsequently shown \cite{Links} that a general
prescription for the construction of integrable systems with periodic
boundary
conditions and quantum algebra invariance existed which could then be
applied to higher spin models such as the spin 1 XXZ Heisenberg chain
\cite{Linkss}.

In the present article we further develop this method by considering the
graded case to derive the Hamiltonian of the anisotropic
supersymmetric $U$ model with quantum supersymmetry on the closed
chain.  We will adopt a nested algebraic Bethe ansatz to solve the model and
this will be presented in detail in Section 3.  Also the energy of the
Hamiltonian will be given.

\section{Quantum algebra invariant Hamiltonian for the supersymmetric
$U$ model}

The following notation will be adopted.  Electrons on a
lattice are described by canonical Fermi operators $c_{i,\sigma}$
and $c_{i,\sigma}^{\dag}$ satisfying the anti-commutation relations
given by $\{c^{\dag}_{i,\sigma},c_{j,\tau}\}=\delta_{ij}
\delta_{\sigma\tau}$, where $i,j=1,2,..,k$ and $\sigma,
\tau=\uparrow,\downarrow.$  The operator $c_{i,\sigma}$
($c_{i,\sigma}^{\dag}$)
annihilates (creates) an electron of spin $\sigma$ at site $i$, which
implies that the Fock vacuum $|0>$ satisfies $c_{i,\sigma}|0>=0$.
At a given lattice site $i$ there are four possible electronic states:
$$|0>,
\mbox{\hspace{.5cm}}|\uparrow>_i=c^{\dag}_{i,\uparrow}|0>,
\mbox{\hspace{.5cm}}|\downarrow>_i=c^{\dag}_{i,\downarrow}|0>,
\mbox{\hspace{.5cm}}|\uparrow\downarrow>_i=c^{\dag}_{i,\downarrow}c^{\dag}_{
i,\uparrow}|0>.$$
By $n_{i,\sigma}=c^{\dag}_{i,\sigma}c_{i,\sigma}$ we denote the number
operator for electrons with spin
$\sigma$ on site $i$, and we write $n_i=n_{i,\uparrow}+n_{i,\downarrow}$.
The local Hamiltonian for this model is \cite{bar}
\begin{eqnarray}
H_{i(i+1)}& =& -\sum_{\sigma} (c_{i\sigma}^{\dag}c_{i+1\sigma}+h.c.)
\mbox{exp}\left[-\frac 12(\zeta-\sigma \gamma)n_{i,-\sigma}-\frac
12(\zeta+\sigma \gamma)
n_{i+1,- \sigma}\right]\nonumber \\
&&+\left[ Un_{i\uparrow}n_{i\downarrow}+Un_{i+1\uparrow}n_{i+1\downarrow} +
U(c^{\dag}_{i\uparrow}c^{\dag}_{i\downarrow}c_{i+1 \downarrow}c_{i+1
\uparrow} +h.c.)\right], \label{ham1}
\end{eqnarray}
where $i$ labels the sites and
$$ U = \epsilon[2 e^{-\zeta} \mbox{cosh } \zeta -
\mbox{cosh }\gamma)]^{\frac 12}, \mbox{\hspace{1cm}} \epsilon=\pm 1.$$
This Hamiltonian may be obtained from the $R$-matrix of a one-parameter
family of four-dimensional representations of $U_q[gl(2|1)]$,
which is afforded by the module $W$ with highest
weight $(0,0|\alpha)$.  The details of this construction may be found
in \cite{me}.

The Hamiltonian (\ref{ham1}) may be modified
to ensure quantum superalgebra invariance by adapting the general
construction
presented in \cite{Links}.
We can write
$$H=\sum_{i=1}^{k-1}H_{i(i+1)} +H_{1k},$$
where the boundary term is given by
$$H_{1k}=G H_{k,1} G^{-1}$$
with
$$G= R^-_{21} R^-_{31} ...R^-_{k1},~~~k\mbox{ the lattice length}.$$
Above, $R^-$ is the constant $R$-matrix obtained as the zero spectral
parameter limit of the Yang-Baxter equation solution
associated with the model \cite{me}. These operators act in the quantum
space and the
closed boundary conditions  of the model may be explained by the relations
$$GH_{i,i+1}=H_{i+1,i+2}G,~~~i=1,2,...,k-2,~~~ GH_{1k}=H_{12}G.$$
The quantum supersymmetry of the Hamiltonian is a result of the
intertwining properties of the matrices $R$.

\section{Nested Algebraic Bethe ansatz}

We present the nested algebraic Bethe ansatz for the above
Hamiltonian by extending the methods presented in \cite{Links,Linkss} to
treat
quantum group invariant closed higher spin chains to the
graded case.  We begin with the $R$-matrix satisfying
the Yang-Baxter equation constructed directly from a solution of the twisted
representation as given in \cite{meqaba}.

The Yang-Baxter Equation may be written as the operator equation:
\begin{eqnarray}
{}_{vv}R^{\alpha_1 \alpha_2}_{\beta_1 \beta_2}(x/y)~~ {}_{vw}R^{\beta_1
a}_{\gamma_1 b}(x)~~
{}_{vw}R^{\beta_2 b}_{\gamma_2 c}(y) = {}_{vw}R^{\alpha_2 a}_{\beta_2
b}(y)~~{}_{vw}R^{\alpha_1 b}_{\beta_1 c}(x)~~
{}_{vv}R^{\beta_1 \beta_2}_{\gamma_1 \gamma_2 }(x/y) ,\label{vvham}
\end{eqnarray}
acting on the spaces $V\otimes V\otimes W$ where $V$ is the vector module
and $W$ is the four-dimensional module associated with the one-parameter
family of minimal typical representations.  Greek
indices are used to label the matrix spaces, that is the first two
spaces and Roman indices label the quantum space, which is the third
space.  The quantum space represents the Hilbert space over a site on the
one-dimensional lattice.  The ${}_{vv}R$-matrix acts in the matrix space
and it is between the two matrix spaces that the graded tensor product
acts.

The ${}_{vv}R$-matrix acts on $V\otimes V$ and has the
form \cite{for}, \cite{vecrep}
\begin{eqnarray}
{}_{vv}R^{\beta_1 \beta_2}_{\alpha_1 \alpha_2}(x)=
\ba{c}
\unitlength=0.50mm
\begin{picture}(20.,25.)
\put(-3.,2.){\makebox(0.,0.){$\s x$}}
\put(0.,-4.){\makebox(0.,0.){$\s \alpha_2 $}}
\put(23.,2.){\makebox(0.,0.){$\s 1$}}
\put(20.,-4.){\makebox(0.,0.){$\s \alpha_1 $}}
\put(0.,24.){\makebox(0.,0.){$\s \beta_1 $}}
\put(20.,24.){\makebox(0.,0.){$\s \beta_2 $}}
\put(0.,20){\vector(-1,1){1.}}
\put(20.,20){\vector(1,1){1.}}
\put(0.00,0.00){\circle{2.0}}
\put(1.00,1.00){\circle{2.0}}
\put(2.00,2.00){\circle{2.0}}
\put(3.00,3.00){\circle{2.0}}
\put(4.00,4.00){\circle{2.0}}
\put(5.00,5.00){\circle{2.0}}
\put(6.00,6.00){\circle{2.0}}
\put(7.00,7.00){\circle{2.0}}
\put(8.00,8.00){\circle{2.0}}
\put(9.00,9.00){\circle{2.0}}
\put(10.00,10.00){\circle{2.0}}
\put(11.00,11.00){\circle{2.0}}
\put(12.00,12.00){\circle{2.0}}
\put(13.00,13.00){\circle{2.0}}
\put(14.00,14.00){\circle{2.0}}
\put(15.00,15.00){\circle{2.0}}
\put(16.00,16.00){\circle{2.0}}
\put(17.00,17.00){\circle{2.0}}
\put(18.00,18.00){\circle{2.0}}
\put(19.00,19.00){\circle{2.0}}
\put(20.00,0.00){\circle{2.0}}
\put(19.00,1.00){\circle{2.0}}
\put(18.00,2.00){\circle{2.0}}
\put(17.00,3.00){\circle{2.0}}
\put(16.00,4.00){\circle{2.0}}
\put(15.00,5.00){\circle{2.0}}
\put(14.00,6.00){\circle{2.0}}
\put(13.00,7.00){\circle{2.0}}
\put(12.00,8.00){\circle{2.0}}
\put(11.00,9.00){\circle{2.0}}
\put(10.00,10.00){\circle{2.0}}
\put(9.00,11.00){\circle{2.0}}
\put(8.00,12.00){\circle{2.0}}
\put(7.00,13.00){\circle{2.0}}
\put(6.00,14.00){\circle{2.0}}
\put(5.00,15.00){\circle{2.0}}
\put(4.00,16.00){\circle{2.0}}
\put(3.00,17.00){\circle{2.0}}
\put(2.00,18.00){\circle{2.0}}
\put(1.00,19.00){\circle{2.0}}
\end{picture}
\ea =
\left(\begin{array}{ccccccccc}
A & 0 & 0 & 0 & 0 & 0 & 0 & 0 & 0\\
0&E&0&C&0&0&0&0&0\\0&0&E&0&0&0&C&0&0\\
0&xC&0&E&0&0&0&0&0\\0&0&0&0&A&0&0&0&0\\
0&0&0&0&0&E&0&C&0\\0&0&xC&0&0&0&E&0&0\\
0&0&0&0&0&xC&0&E&0\\0&0&0&0&0&0&0&0&1
\end{array}\right)\nonumber
\end{eqnarray}
where $A,~E$ and $C$ depend on the spectral parameter as follows
$A(x)= \frac {1-xq^2}{x-q^2}$,
$E(x)= \frac {(1-x)q}{x-q^2}$ and $C(x)=\frac{1-q^2}{x-q^2}$.  The
${}_{vv}R$-matrix satisfies the Yang-Baxter
equation
$$R_{12}(x/y) R_{13}(x) R_{23}(y) = R_{23}(y) R_{13}(x)
R_{12}(x/y).\label{ybe}$$
By construction, these $R$-matrices also satisfy the generalized
Cherednik reflection property \cite{22}
\begin{eqnarray}
R_{\alpha'\beta'}^{\alpha\beta}(x)R^{-1}{}_{\gamma\delta}^{\alpha'\beta'}(1/
y)
=R_{\alpha'\beta'}^{\alpha\beta}(y)R^{-1}{}^{\alpha'\beta'}_{\gamma
\delta}(1/x),\label{refl}
\end{eqnarray}
and crossing unitarity \cite{23}
\begin{eqnarray}
R^{st_1}{}_{\alpha' \beta'}^{\alpha\beta}(x\zeta) K^{\alpha'}_{\alpha''}
(R^{-1}){}^{st_1}{}_{\gamma'\delta}^{\alpha''\beta'}(x)
K^{-1} {}^{\gamma'}_{\gamma}=\delta^{\alpha}_{\gamma}
\delta^{\beta}_{\delta},\label{unit}
\end{eqnarray}
where $st_1$ denotes matrix supertransposition in the first space and
$K$ is the crossing matrix given below.

It will be necessary to rewrite the ${}_{vv}R$-matrix in terms of
constant matrices ${}_{vv}R^+$ and ${}_{vv}R^-$ that is,
$${}_{vv} R(x) =\left(\frac {-1}{x-q^2}\right) (x~~{}_{vv}R^+ -
~~{}_{vv}R^- ),$$
where ${}_{vv}R^+$ (${}_{vv}R^-$) corresponds to the leading term in the
limit as $x\rightarrow\infty$ ($x\rightarrow 0$).

The ${}_{vw}R$-matrix was constructed in \cite{meqaba}
in  the $V\otimes W$ space
and has the following form

\begin{eqnarray}
{}_{vw}R^{\beta j}_{\alpha i}(x)=
\ba{c}
\unitlength=0.50mm

\begin{picture}(20.,25.)
\put(-3.,2.){\makebox(0.,0.){$\s x$}}
\put(0.,-4.){\makebox(0.,0.){$\s i$}}
\put(23.,2.){\makebox(0.,0.){$\s 1$}}
\put(20.,-4.){\makebox(0.,0.){$\s \alpha $}}
\put(0.,24.){\makebox(0.,0.){$\s \beta $}}
\put(20.,24.){\makebox(0.,0.){$\s j $}}
\put(0.,20){\vector(-1,1){1.}}
\put(20.,20){\vector(1,1){1.}}
\put(20.00,0.00){\circle{2.0}}
\put(19.00,1.00){\circle{2.0}}
\put(18.00,2.00){\circle{2.0}}
\put(17.00,3.00){\circle{2.0}}
\put(16.00,4.00){\circle{2.0}}
\put(15.00,5.00){\circle{2.0}}
\put(14.00,6.00){\circle{2.0}}
\put(13.00,7.00){\circle{2.0}}
\put(12.00,8.00){\circle{2.0}}
\put(11.00,9.00){\circle{2.0}}
\put(10.00,10.00){\circle{2.0}}
\put(9.00,11.00){\circle{2.0}}
\put(8.00,12.00){\circle{2.0}}
\put(7.00,13.00){\circle{2.0}}
\put(6.00,14.00){\circle{2.0}}
\put(5.00,15.00){\circle{2.0}}
\put(4.00,16.00){\circle{2.0}}
\put(3.00,17.00){\circle{2.0}}
\put(2.00,18.00){\circle{2.0}}
\put(1.00,19.00){\circle{2.0}}
\put(0.,0.){\line(1,1){20.}}
\end{picture}
\ea =
\left(\begin{array}{cccccccccccc}
J&0&0&0&0&0&0&0& 0 & 0 & 0 & 0\\
0&J&0&0&0&0&0&0& 0 & 0 & 0 & 0\\
0&0&Y&0&0&Q'&0&0& -S' & 0 & 0 & 0\\
0&0&0&L&0&0&0&0& 0 & -T' & 0 & 0\\
0&0&0&0&J&0&0&0& 0 & 0 & 0 & 0\\
0&0&Q&0&0&Y&0&0& -P' & 0 & 0 & 0\\
0&0&0&0&0&0&J&0& 0 & 0 & 0 & 0\\
0&0&0&0&0&0&0&L& 0 & 0 & -T' & 0\\
0&0&S&0&0&P&0&0& M & 0 & 0 & 0\\
0&0&0&-T&0&0&0&0& 0 & -N & 0 & 0\\
0&0&0&0&0&0&0&-T& 0 & 0 & -N & 0\\
0&0&0&0&0&0&0&0& 0 & 0 & 0 & 1
\end{array}\right)\nonumber
\end{eqnarray}
where the dependence of these elements on the spectral parameter is given by
\begin{eqnarray}
J(x)&=& \frac
{(x-q^{-\alpha-2})}{(xq^{-\alpha-2}-1 )},~~~Y(x)=J(x)(D+B)+\frac
1{[\alpha+2]},\nonumber \\
L(x)&=& \frac 1{[\alpha+2]} ([\alpha+1] J(x)+1),~~~M(x)=F^2 DJ(x) +\frac
{[\alpha]}{[\alpha+2]},~~~ N(x)= \frac 1{[\alpha+2]} (J(x) +[\alpha+1]), \nonumber \\
Q(x)&=&(qB-Dq^{-1})J(x) -\frac {q^{-1}}{[\alpha+2]},~~~Q'(x)=(q^{-1}
B-qD)J(x) -\frac q{[\alpha+2]},\nonumber \\
S(x)&=& \frac{\sqrt{[\alpha]} }{[\alpha+2]}
q^{-(\alpha+3)/2} -q^{(\alpha+1)/2}F D J(x), ~~~
S'(x)= -\frac{\sqrt{[\alpha]} }{[\alpha+2]}q^{(\alpha+3)/2}
+q^{-(\alpha+1)/2}F D J(x),\nonumber \\
T(x)&=&
\frac{\sqrt{[\alpha+1]}}{[\alpha+2]}(q^{(\alpha+2)/2}J(x) -q^{-(\alpha+2)/2}
 ), ~~~
T'(x)=
\frac{\sqrt{[\alpha+1]}}{[\alpha+2]}(q^{(\alpha+2)/2} -q^{-(\alpha+2)/2}J(x)
 ),\nonumber \\
P(x)&=& q^{(\alpha+3)/2} F D J(x) - \frac {\sqrt{[\alpha]} }
{[\alpha+2]}q^{-(\alpha+1)/2} , ~~~P'(x) = -q^{-(\alpha+3)/2} F DJ(x) + \frac {\sqrt{[\alpha]} }
{[\alpha+2]}q^{(\alpha+1)/2} ,\nonumber 
\end{eqnarray}
with constants
\begin{eqnarray}
D&=& \frac {[\alpha]}{[\alpha+2](q+q^{-1})},
~~~F=\frac{(q+q^{-1})}{\sqrt{[\alpha]} }, ~~~
B= 1/(q+q^{-1}), ~~  \mbox{ and}~~~
[\alpha]=\frac {q^{\alpha} - q^{-\alpha}}{q-q^{-1} }.\nonumber
\end{eqnarray}
The ${}_{vw}R$-matrix as well as satisfying the Yang Baxter relation
(\ref{vvham}), also satisfies the generalized
Cherednick reflection property (\ref{refl}) and crossing unitarity
(\ref{unit}).

We now introduce an auxiliary doubled monodromy matrix
\begin{eqnarray}
\nonumber
\lefteqn{{}_{vw}U(x)^{\beta \{j\}}_{\alpha \{i\} } =
\ba{c}
\unitlength=0.50mm
\begin{picture}(95.,49.)
\put(45.,15.){\makebox(0,0)[cc]{$\cdots$}}
\put(21.,-3.){\makebox(0,0)[cc]{$\s i_1$}}
\put(31.,-3.){\makebox(0,0)[cc]{$\s i_2$}}
\put(71.,-3.){\makebox(0,0)[cc]{$\s i_k$}}
\put(21.,47.){\makebox(0,0)[cc]{$\s j_1$}}
\put(31.,47.){\makebox(0,0)[cc]{$\s j_2$}}
\put(71.,47.){\makebox(0,0)[cc]{$\s j_k$}}
\put(86.,15.){\makebox(0,0)[cc]{$\s \alpha$}}
\put(70.,0.){\vector(0,1){45.}}
\put(30.,0.){\vector(0,1){45.}}
\put(20.,0.){\vector(0,1){45.}}
\put(80.,35.){\vector(1,0){1.}}
\put(57.00,35.00){\circle{2.}}
\put(58.00,35.00){\circle{2.}}
\put(59.00,35.00){\circle{2.}}
\put(60.00,35.00){\circle{2.}}
\put(61.00,35.00){\circle{2.}}
\put(62.00,35.00){\circle{2.}}
\put(63.00,35.00){\circle{2.}}
\put(64.00,35.00){\circle{2.}}
\put(65.00,35.00){\circle{2.}}
\put(66.00,35.00){\circle{2.}}
\put(67.00,35.00){\circle{2.}}
\put(68.00,35.00){\circle{2.}}
\put(72.00,35.00){\circle{2.}}
\put(73.00,35.00){\circle{2.}}
\put(74.00,35.00){\circle{2.}}
\put(75.00,35.00){\circle{2.}}
\put(76.00,35.00){\circle{2.}}
\put(77.00,35.00){\circle{2.}}
\put(78.00,35.00){\circle{2.}}
\put(57.00,15.00){\circle{2.}}
\put(58.00,15.00){\circle{2.}}
\put(59.00,15.00){\circle{2.}}
\put(60.00,15.00){\circle{2.}}
\put(61.00,15.00){\circle{2.}}
\put(62.00,15.00){\circle{2.}}
\put(63.00,15.00){\circle{2.}}
\put(64.00,15.00){\circle{2.}}
\put(65.00,15.00){\circle{2.}}
\put(66.00,15.00){\circle{2.}}
\put(67.00,15.00){\circle{2.}}
\put(68.00,15.00){\circle{2.}}
\put(69.00,15.00){\circle{2.}}
\put(70.00,15.00){\circle{2.}}
\put(71.00,15.00){\circle{2.}}
\put(72.00,15.00){\circle{2.}}
\put(73.00,15.00){\circle{2.}}
\put(74.00,15.00){\circle{2.}}
\put(75.00,15.00){\circle{2.}}
\put(76.00,15.00){\circle{2.}}
\put(77.00,15.00){\circle{2.}}
\put(78.00,15.00){\circle{2.}}
\put(79.00,15.00){\circle{2.}}
\put(80.00,15.00){\circle{2.}}
\put(45.,35.){\makebox(0,0)[cc]{$\cdots$}}
\put(86.,35.){\makebox(0,0)[cc]{$\s \beta$}}
\put(15.00,15.00){\circle{2.}}
\put(16.00,15.00){\circle{2.}}
\put(17.00,15.00){\circle{2.}}
\put(18.00,15.00){\circle{2.}}
\put(19.00,15.00){\circle{2.}}
\put(20.00,15.00){\circle{2.}}
\put(21.00,15.00){\circle{2.}}
\put(22.00,15.00){\circle{2.}}
\put(23.00,15.00){\circle{2.}}
\put(24.00,15.00){\circle{2.}}
\put(25.00,15.00){\circle{2.}}
\put(26.00,15.00){\circle{2.}}
\put(27.00,15.00){\circle{2.}}
\put(28.00,15.00){\circle{2.}}
\put(29.00,15.00){\circle{2.}}
\put(30.00,15.00){\circle{2.}}
\put(31.00,15.00){\circle{2.}}
\put(32.00,15.00){\circle{2.}}
\put(33.00,15.00){\circle{2.}}
\put(34.00,15.00){\circle{2.}}
\put(35.00,15.00){\circle{2.}}
\put(36.00,15.00){\circle{2.}}
\put(37.00,15.00){\circle{2.}}
\put(15.00,35.00){\circle{2.}}
\put(16.00,35.00){\circle{2.}}
\put(17.00,35.00){\circle{2.}}
\put(18.00,35.00){\circle{2.}}
\put(22.00,35.00){\circle{2.}}
\put(23.00,35.00){\circle{2.}}
\put(24.00,35.00){\circle{2.}}
\put(25.00,35.00){\circle{2.}}
\put(26.00,35.00){\circle{2.}}
\put(28.00,35.){\vector(1,0){1.}}
\put(32.00,35.00){\circle{2.}}
\put(33.00,35.00){\circle{2.}}
\put(34.00,35.00){\circle{2.}}
\put(35.00,35.00){\circle{2.}}
\put(36.00,35.00){\circle{2.}}
\put(37.00,35.00){\circle{2.}}
\put(39.00,35.){\vector(1,0){1.}}
\put(5.00,25.){\circle*{2.5}}
\put(5.00,25.00){\circle{2.}}
\put(6.00,26.00){\circle{2.}}
\put(7.00,27.00){\circle{2.}}
\put(8.00,28.00){\circle{2.}}
\put(9.00,29.00){\circle{2.}}
\put(10.00,30.00){\circle{2.}}
\put(11.00,31.00){\circle{2.}}
\put(12.00,32.00){\circle{2.}}
\put(13.00,33.00){\circle{2.}}
\put(14.00,34.00){\circle{2.}}
\put(15.00,35.00){\circle{2.}}
\put(5.00,25.00){\circle{2.}}
\put(6.00,24.00){\circle{2.}}
\put(7.00,23.00){\circle{2.}}
\put(8.00,22.00){\circle{2.}}
\put(9.00,21.00){\circle{2.}}
\put(10.00,20.00){\circle{2.}}
\put(11.00,19.00){\circle{2.}}
\put(12.00,18.00){\circle{2.}}
\put(13.00,17.00){\circle{2.}}
\put(14.00,16.00){\circle{2.}}
\put(15.00,15.00){\circle{2.}}
\end{picture}
\ea}\\\nonumber
&=&
{}_{vw}R_{+}{}^{\beta_2 j_1}_{\alpha' j'_1}~
{}_{vw}R_{+}{}^{\beta_3 j_2}_{\beta_2 j'_2} \dots
{}_{vw}R_{+}{}^{\beta j_k}_{\beta_k j'_k}~
{}_{vw}R_{\alpha_2 i_1}^{\alpha' j'_1}(1/x)~
{}_{vw}R_{\alpha_3 i_2}^{\alpha_2 j_2'}(1/x)\dots
{}_{vw}R_{\alpha i_l}^{\alpha_kj'_k}(1/x),\label{mono}
\end{eqnarray}
acting on $V \otimes W^{\otimes k}$.
Above ${}_{vw}R_+ $ represents the leading term in the matrix
${}_{vw}R^{-1}(x)$ for the limit as $x\rightarrow \infty$.

Represent the doubled monodromy matrix in the following way:
\begin{eqnarray}
{}_{vw}U^\gamma_\alpha(x)&=& \left( \begin{array}{clcr}
{}_{vw}U_1^1(x) & {}_{vw}U_2^1(x) & {}_{vw}U_3^1(x)\\
{}_{vw}U_1^2(x) & {}_{vw}U_2^2(x) & {}_{vw}U_2^3(x) \\
{}_{vw}U_1^3(x) & {}_{vw}U_2^3(x) & {}_{vw}U_3^3(x)
\end{array}\right).\label{umatrix}
\end{eqnarray}
It may be shown that this monodromy matrix satisfies the following
Yang-Baxter relation
\begin{eqnarray}
{}_{vv}R^{\alpha_1 \beta_1}_{\alpha_2 \beta_2}(y/x)
~~{}_{vw}U^{\beta_2 a}_{\gamma_2 b}(x)
~~{}_{vv}R_+{}^{\gamma_2 \alpha_2}_{\gamma_1 \delta_2}
~~ {}_{vw}U^{\delta_2 b}_{\delta_1 c}(y) = 
{}_{vw}U^{\alpha_1 a}_{\alpha_2 b}(y)
~~{}_{vv}R_+{}^{\alpha_2 \beta_1}_{\delta_2 \beta_2}
~~{}_{vw}U^{\beta_2b}_{\gamma_2 c}(x)
~~{}_{vv}R^{\gamma_2 \delta_2}_{\gamma_1 \delta_1 }(y/x) ,
\label{newybe}
\end{eqnarray}
depicted graphically below.
\[
\unitlength=0.50mm
\begin{picture}(170.,85.)
\put(85.,45.){\makebox(0,0)[cc]{$=$}}
\put(5.,10.){\vector(0,1){75.}}
\put(36.,10.){\vector(0,1){75.}}
\put(15.,10.){\vector(0,1){75.}}
\put(115.,10.){\vector(0,1){75.}}
\put(146.,10.){\vector(0,1){75.}}
\put(125.,10.){\vector(0,1){75.}}
\put(59.,80.){\vector(1,0){1.}}
\put(65.,80.){\makebox(0,0)[cc]{$\s \alpha_1 $}}
\put(65.,70.){\makebox(0,0)[cc]{$\s \beta_1 $}}
\put(65.,50.){\makebox(0,0)[cc]{$\s \gamma_1 $}}
\put(65.,20.){\makebox(0,0)[cc]{$\s \delta_1 $}}
\put(170.,70.){\makebox(0,0)[cc]{$\s \alpha_1 $}}
\put(170.,40.){\makebox(0,0)[cc]{$\s \beta_1 $}}
\put(170.,20.){\makebox(0,0)[cc]{$\s \gamma_1 $}}
\put(170.,10.){\makebox(0,0)[cc]{$\s \delta_1 $}}

\put(-10.00,60.00){\circle*{2.}}
\put(-9.00,61.00){\circle{2.}}
\put(-8.00,62.00){\circle{2.}}
\put(-7.00,63.00){\circle{2.}}
\put(-6.00,64.00){\circle{2.}}
\put(-5.00,65.00){\circle{2.}}
\put(-4.00,66.00){\circle{2.}}
\put(-3.00,67.00){\circle{2.}}
\put(-2.00,68.00){\circle{2.}}
\put(-1.00,69.00){\circle{2.}}
\put(0.00,70.00){\circle{2.}}
\put(-10.00,59.00){\circle*{2.}}
\put(-9.00,58.00){\circle{2.}}
\put(-8.00,57.00){\circle{2.}}
\put(-7.00,56.00){\circle{2.}}
\put(-6.00,55.00){\circle{2.}}
\put(-5.00,54.00){\circle{2.}}
\put(-4.00,53.00){\circle{2.}}
\put(-3.00,52.00){\circle{2.}}
\put(-2.00,51.00){\circle{2.}}
\put(-1.00,50.00){\circle{2.}}

\put(58.,70.){\vector(1,0){2.}}
\put(0.00,70.00){\circle{2.}}
\put(1.00,70.00){\circle{2.}}
\put(2.00,70.00){\circle{2.}}
\put(3.00,70.00){\circle{2.}}
\put(7.00,70.00){\circle{2.}}
\put(8.00,70.00){\circle{2.}}
\put(9.00,70.00){\circle{2.}}
\put(10.00,70.00){\circle{2.}}
\put(11.,70.){\vector(1,0){2.}}
\put(17.00,70.00){\circle{2.}}
\put(18.00,70.00){\circle{2.}}
\put(19.00,70.00){\circle{2.}}
\put(20.00,70.00){\circle{2.}}
\put(21.,70.){\vector(1,0){2.}}
\put(25.,70.){\makebox(0,0)[cc]{$\s \ldots$}}
\put(30.00,70.00){\circle{2.}}
\put(31.00,70.00){\circle{2.}}
\put(32.00,70.00){\circle{2.}}
\put(33.00,70.00){\circle{2.}}
\put(34.00,70.00){\circle{2.}}
\put(39.00,70.00){\circle{2.}}
\put(40.00,70.00){\circle{2.}}
\put(41.00,70.00){\circle{2.}}
\put(42.00,70.00){\circle{2.}}
\put(43.00,70.00){\circle{2.}}
\put(44.00,70.00){\circle{2.}}
\put(46.00,70.00){\circle{2.}}
\put(47.00,70.00){\circle{2.}}
\put(48.00,70.00){\circle{2.}}
\put(49.00,70.00){\circle{2.}}
\put(50.00,70.00){\circle{2.}}
\put(51.00,70.00){\circle{2.}}
\put(52.00,70.00){\circle{2.}}
\put(53.00,70.00){\circle{2.}}
\put(54.00,70.00){\circle{2.}}
\put(55.00,70.00){\circle{2.}}
\put(56.00,70.00){\circle{2.}}
\put(57.00,70.00){\circle{2.}}

\put(0.00,50.00){\circle{2.}}
\put(1.00,50.00){\circle{2.}}
\put(2.00,50.00){\circle{2.}}
\put(3.00,50.00){\circle{2.}}
\put(4.00,50.00){\circle{2.}}
\put(5.00,50.00){\circle{2.}}
\put(6.00,50.00){\circle{2.}}
\put(7.00,50.00){\circle{2.}}
\put(8.00,50.00){\circle{2.}}
\put(9.00,50.00){\circle{2.}}
\put(10.00,50.00){\circle{2.}}
\put(11.00,50.00){\circle{2.}}
\put(12.00,50.00){\circle{2.}}
\put(13.00,50.00){\circle{2.}}
\put(14.00,50.00){\circle{2.}}
\put(15.00,50.00){\circle{2.}}
\put(16.00,50.00){\circle{2.}}
\put(17.00,50.00){\circle{2.}}
\put(18.00,50.00){\circle{2.}}
\put(19.00,50.00){\circle{2.}}
\put(20.00,50.00){\circle{2.}}
\put(21.00,50.00){\circle{2.}}
\put(22.00,50.00){\circle{2.}}
\put(23.00,50.00){\circle{2.}}
\put(24.00,50.00){\circle{2.}}
\put(25.00,50.00){\circle{2.}}
\put(26.00,50.00){\circle{2.}}
\put(27.00,50.00){\circle{2.}}
\put(28.00,50.00){\circle{2.}}
\put(29.00,50.00){\circle{2.}}

\put(30.00,50.00){\circle{2.}}
\put(31.00,50.00){\circle{2.}}
\put(32.00,50.00){\circle{2.}}
\put(33.00,50.00){\circle{2.}}
\put(34.00,50.00){\circle{2.}}
\put(35.00,50.00){\circle{2.}}
\put(36.00,50.00){\circle{2.}}
\put(37.00,50.00){\circle{2.}}
\put(38.00,50.00){\circle{2.}}
\put(39.00,50.00){\circle{2.}}
\put(40.00,50.00){\circle{2.}}
\put(41.00,50.00){\circle{2.}}
\put(42.00,50.00){\circle{2.}}
\put(43.00,50.00){\circle{2.}}
\put(44.00,50.00){\circle{2.}}
\put(45.00,50.00){\circle{2.}}
\put(46.00,50.00){\circle{2.}}
\put(47.00,50.00){\circle{2.}}
\put(48.00,50.00){\circle{2.}}
\put(49.00,50.00){\circle{2.}}
\put(50.00,50.00){\circle{2.}}
\put(51.00,50.00){\circle{2.}}
\put(52.00,50.00){\circle{2.}}
\put(53.00,50.00){\circle{2.}}
\put(54.00,50.00){\circle{2.}}
\put(55.00,50.00){\circle{2.}}
\put(56.00,50.00){\circle{2.}}
\put(57.00,50.00){\circle{2.}}
\put(58.00,50.00){\circle{2.}}
\put(59.00,50.00){\circle{2.}}

\put(110.00,40.00){\circle{2.}}
\put(111.00,40.00){\circle{2.}}
\put(112.00,40.00){\circle{2.}}
\put(123.,40.){\vector(1,0){2.}}
\put(117.00,40.00){\circle{2.}}
\put(118.00,40.00){\circle{2.}}
\put(119.00,40.00){\circle{2.}}
\put(120.00,40.00){\circle{2.}}
\put(121.00,40.00){\circle{2.}}
\put(122.00,40.00){\circle{2.}}
\put(131.,40.){\vector(1,0){2.}}
\put(127.00,40.00){\circle{2.}}
\put(128.00,40.00){\circle{2.}}
\put(129.00,40.00){\circle{2.}}
\put(130.00,40.00){\circle{2.}}
\put(142.00,40.00){\circle{2.}}
\put(143.00,40.00){\circle{2.}}
\put(144.00,40.00){\circle{2.}}
\put(148.00,40.00){\circle{2.}}
\put(149.00,40.00){\circle{2.}}
\put(150.00,40.00){\circle{2.}}
\put(151.00,40.00){\circle{2.}}
\put(157.00,40.00){\circle{2.}}
\put(158.00,40.00){\circle{2.}}
\put(159.00,40.00){\circle{2.}}
\put(160.00,40.00){\circle{2.}}
\put(161.00,40.00){\circle{2.}}
\put(162.00,40.00){\circle{2.}}
\put(163.00,40.00){\circle{2.}}
\put(164.00,40.00){\circle{2.}}

\put(157.,40.){\vector(1,0){10.}}
\put(110.00,20.00){\circle{2.}}
\put(111.00,20.00){\circle{2.}}
\put(112.00,20.00){\circle{2.}}
\put(113.00,20.00){\circle{2.}}
\put(114.00,20.00){\circle{2.}}
\put(115.00,20.00){\circle{2.}}
\put(116.00,20.00){\circle{2.}}
\put(117.00,20.00){\circle{2.}}
\put(118.00,20.00){\circle{2.}}
\put(119.00,20.00){\circle{2.}}
\put(120.00,20.00){\circle{2.}}
\put(121.00,20.00){\circle{2.}}
\put(122.00,20.00){\circle{2.}}
\put(123.00,20.00){\circle{2.}}
\put(124.00,20.00){\circle{2.}}
\put(125.00,20.00){\circle{2.}}
\put(126.00,20.00){\circle{2.}}
\put(127.00,20.00){\circle{2.}}
\put(128.00,20.00){\circle{2.}}
\put(129.00,20.00){\circle{2.}}
\put(130.00,20.00){\circle{2.}}
\put(142.00,20.00){\circle{2.}}
\put(143.00,20.00){\circle{2.}}
\put(144.00,20.00){\circle{2.}}
\put(145.00,20.00){\circle{2.}}
\put(146.00,20.00){\circle{2.}}
\put(147.00,20.00){\circle{2.}}
\put(148.00,20.00){\circle{2.}}
\put(149.00,20.00){\circle{2.}}
\put(150.00,20.00){\circle{2.}}
\put(151.00,20.00){\circle{2.}}
\put(152.00,20.00){\circle{2.}}
\put(153.00,20.00){\circle{2.}}
\put(154.00,20.00){\circle{2.}}
\put(155.00,20.00){\circle{2.}}
\put(156.00,20.00){\circle{2.}}
\put(157.00,20.00){\circle{2.}}
\put(158.00,20.00){\circle{2.}}
\put(159.00,20.00){\circle{2.}}
\put(160.00,20.00){\circle{2.}}
\put(161.00,20.00){\circle{2.}}
\put(162.00,20.00){\circle{2.}}
\put(163.00,20.00){\circle{2.}}
\put(164.00,20.00){\circle{2.}}
\put(165.00,20.00){\circle{2.}}
\put(166.00,20.00){\circle{2.}}

\put(25.,50.){\makebox(0,0)[cc]{$\s \ldots$}}
\put(136.,40.){\makebox(0,0)[cc]{$\s \ldots$}}
\put(136.,20.){\makebox(0,0)[cc]{$\s \ldots$}}

\put(100.00,30.00){\circle*{2.3}}
\put(101.00,31.00){\circle{2.}}
\put(102.00,32.00){\circle{2.}}
\put(103.00,33.00){\circle{2.}}
\put(104.00,34.00){\circle{2.}}
\put(105.00,35.00){\circle{2.}}
\put(106.00,36.00){\circle{2.}}
\put(107.00,37.00){\circle{2.}}
\put(108.00,38.00){\circle{2.}}
\put(109.00,39.00){\circle{2.}}
\put(110.00,40.00){\circle{2.}}

\put(100.00,30.00){\circle{2.}}
\put(101.00,29.00){\circle{2.}}
\put(102.00,28.00){\circle{2.}}
\put(103.00,27.00){\circle{2.}}
\put(104.00,26.00){\circle{2.}}
\put(105.00,25.00){\circle{2.}}
\put(106.00,24.00){\circle{2.}}
\put(107.00,23.00){\circle{2.}}
\put(108.00,22.00){\circle{2.}}
\put(109.00,21.00){\circle{2.}}
\put(110.00,20.00){\circle{2.}}

\put(-10.00,30.00){\circle*{2.3}}
\put(-9.00,31.00){\circle{2.}}
\put(-8.00,32.00){\circle{2.}}
\put(-7.00,33.00){\circle{2.}}
\put(-6.00,34.00){\circle{2.}}
\put(-5.00,35.00){\circle{2.}}
\put(-4.00,36.00){\circle{2.}}
\put(-3.00,37.00){\circle{2.}}
\put(-2.00,38.00){\circle{2.}}
\put(-1.00,39.00){\circle{2.}}
\put(0.00,40.00){\circle{2.}}

\put(-10.00,30.00){\circle{2.}}
\put(-9.00,29.00){\circle{2.}}
\put(-8.00,28.00){\circle{2.}}
\put(-7.00,27.00){\circle{2.}}
\put(-6.00,26.00){\circle{2.}}
\put(-5.00,25.00){\circle{2.}}
\put(-4.00,24.00){\circle{2.}}
\put(-3.00,23.00){\circle{2.}}
\put(-2.00,22.00){\circle{2.}}
\put(-1.00,21.00){\circle{2.}}
\put(0.00,20.00){\circle{2.}}

\put(100.00,60.00){\circle*{2.3}}
\put(101.00,61.00){\circle{2.}}
\put(102.00,62.00){\circle{2.}}
\put(103.00,63.00){\circle{2.}}
\put(104.00,64.00){\circle{2.}}
\put(105.00,65.00){\circle{2.}}
\put(106.00,66.00){\circle{2.}}
\put(107.00,67.00){\circle{2.}}
\put(108.00,68.00){\circle{2.}}
\put(109.00,69.00){\circle{2.}}
\put(110.00,70.00){\circle{2.}}

\put(100.00,60.00){\circle{2.}}
\put(101.00,59.00){\circle{2.}}
\put(102.00,58.00){\circle{2.}}
\put(103.00,57.00){\circle{2.}}
\put(104.00,56.00){\circle{2.}}
\put(105.00,55.00){\circle{2.}}
\put(106.00,54.00){\circle{2.}}
\put(107.00,53.00){\circle{2.}}
\put(108.00,52.00){\circle{2.}}
\put(109.00,51.00){\circle{2.}}
\put(110.00,50.00){\circle{2.}}

\put(111.00,50.00){\circle{2.}}
\put(112.00,50.00){\circle{2.}}
\put(113.00,50.00){\circle{2.}}
\put(114.00,50.00){\circle{2.}}
\put(115.00,50.00){\circle{2.}}
\put(116.00,50.00){\circle{2.}}
\put(117.00,50.00){\circle{2.}}
\put(118.00,50.00){\circle{2.}}
\put(119.00,50.00){\circle{2.}}
\put(120.00,50.00){\circle{2.}}
\put(121.00,50.00){\circle{2.}}
\put(122.00,50.00){\circle{2.}}
\put(123.00,50.00){\circle{2.}}
\put(124.00,50.00){\circle{2.}}
\put(125.00,50.00){\circle{2.}}
\put(126.00,50.00){\circle{2.}}
\put(127.00,50.00){\circle{2.}}
\put(128.00,50.00){\circle{2.}}
\put(129.00,50.00){\circle{2.}}
\put(130.00,50.00){\circle{2.}}
\put(131.00,50.00){\circle{2.}}
\put(136.,50.){\makebox(0,0)[cc]{$\s \ldots$}}
\put(142.00,50.00){\circle{2.}}
\put(143.00,50.00){\circle{2.}}
\put(144.00,50.00){\circle{2.}}
\put(145.00,50.00){\circle{2.}}
\put(146.00,50.00){\circle{2.}}
\put(147.00,50.00){\circle{2.}}
\put(148.00,50.00){\circle{2.}}
\put(149.00,50.00){\circle{2.}}
\put(150.00,50.00){\circle{2.}}
\put(151.00,50.00){\circle{2.}}
\put(152.00,50.00){\circle{2.}}
\put(153.00,50.00){\circle{2.}}
\put(154.00,50.00){\circle{2.}}
\put(154.00,49.00){\circle{2.}}
\put(154.00,48.00){\circle{2.}}
\put(154.00,47.00){\circle{2.}}
\put(154.00,46.00){\circle{2.}}
\put(154.00,45.00){\circle{2.}}
\put(154.00,44.00){\circle{2.}}
\put(154.00,43.00){\circle{2.}}
\put(154.00,42.00){\circle{2.}}
\put(154.00,41.00){\circle{2.}}
\put(154.00,40.00){\circle{2.}}
\put(154.00,39.00){\circle{2.}}
\put(154.00,38.00){\circle{2.}}
\put(154.00,37.00){\circle{2.}}
\put(154.00,36.00){\circle{2.}}
\put(154.00,35.00){\circle{2.}}
\put(154.00,34.00){\circle{2.}}
\put(154.00,33.00){\circle{2.}}
\put(154.00,32.00){\circle{2.}}
\put(154.00,31.00){\circle{2.}}
\put(154.00,30.00){\circle{2.}}
\put(154.00,29.00){\circle{2.}}
\put(154.00,28.00){\circle{2.}}
\put(154.00,27.00){\circle{2.}}
\put(154.00,26.00){\circle{2.}}
\put(154.00,25.00){\circle{2.}}
\put(154.00,24.00){\circle{2.}}
\put(154.00,23.00){\circle{2.}}
\put(154.00,22.00){\circle{2.}}
\put(154.00,21.00){\circle{2.}}
\put(154.00,20.00){\circle{2.}}
\put(154.00,19.00){\circle{2.}}
\put(154.00,18.00){\circle{2.}}
\put(154.00,17.00){\circle{2.}}
\put(154.00,16.00){\circle{2.}}
\put(154.00,15.00){\circle{2.}}
\put(154.00,14.00){\circle{2.}}
\put(154.00,13.00){\circle{2.}}
\put(154.00,12.00){\circle{2.}}
\put(154.00,11.00){\circle{2.}}
\put(154.00,10.00){\circle{2.}}
\put(155.00,10.00){\circle{2.}}
\put(156.00,10.00){\circle{2.}}
\put(157.00,10.00){\circle{2.}}
\put(158.00,10.00){\circle{2.}}
\put(159.00,10.00){\circle{2.}}
\put(160.00,10.00){\circle{2.}}
\put(161.00,10.00){\circle{2.}}
\put(162.00,10.00){\circle{2.}}
\put(163.00,10.00){\circle{2.}}
\put(164.00,10.00){\circle{2.}}
\put(165.00,10.00){\circle{2.}}
\put(166.00,10.00){\circle{2.}}

\put(1.00,20.00){\circle{2.}}
\put(2.00,20.00){\circle{2.}}
\put(3.00,20.00){\circle{2.}}
\put(4.00,20.00){\circle{2.}}
\put(5.00,20.00){\circle{2.}}
\put(6.00,20.00){\circle{2.}}
\put(7.00,20.00){\circle{2.}}
\put(8.00,20.00){\circle{2.}}
\put(9.00,20.00){\circle{2.}}
\put(10.00,20.00){\circle{2.}}
\put(11.00,20.00){\circle{2.}}
\put(12.00,20.00){\circle{2.}}
\put(13.00,20.00){\circle{2.}}
\put(14.00,20.00){\circle{2.}}
\put(15.00,20.00){\circle{2.}}
\put(16.00,20.00){\circle{2.}}
\put(17.00,20.00){\circle{2.}}
\put(18.00,20.00){\circle{2.}}
\put(19.00,20.00){\circle{2.}}
\put(25.,20.){\makebox(0,0)[cc]{$\s \ldots$}}
\put(31.00,20.00){\circle{2.}}
\put(32.00,20.00){\circle{2.}}
\put(33.00,20.00){\circle{2.}}
\put(34.00,20.00){\circle{2.}}
\put(35.00,20.00){\circle{2.}}
\put(36.00,20.00){\circle{2.}}
\put(37.00,20.00){\circle{2.}}
\put(38.00,20.00){\circle{2.}}
\put(39.00,20.00){\circle{2.}}
\put(40.00,20.00){\circle{2.}}
\put(41.00,20.00){\circle{2.}}
\put(42.00,20.00){\circle{2.}}
\put(43.00,20.00){\circle{2.}}
\put(44.00,20.00){\circle{2.}}
\put(45.00,20.00){\circle{2.}}
\put(46.00,20.00){\circle{2.}}
\put(47.00,20.00){\circle{2.}}
\put(48.00,20.00){\circle{2.}}
\put(49.00,20.00){\circle{2.}}
\put(50.00,20.00){\circle{2.}}
\put(51.00,20.00){\circle{2.}}
\put(52.00,20.00){\circle{2.}}
\put(53.00,20.00){\circle{2.}}
\put(54.00,20.00){\circle{2.}}
\put(55.00,20.00){\circle{2.}}
\put(56.00,20.00){\circle{2.}}
\put(57.00,20.00){\circle{2.}}
\put(58.00,20.00){\circle{2.}}
\put(59.00,20.00){\circle{2.}}
\put(60.00,20.00){\circle{2.}}
\put(61.00,20.00){\circle{2.}}

\put(1.00,40.00){\circle{2.}}
\put(2.00,40.00){\circle{2.}}
\put(3.00,40.00){\circle{2.}}
\put(7.00,40.00){\circle{2.}}
\put(8.00,40.00){\circle{2.}}
\put(9.00,40.00){\circle{2.}}
\put(10.00,40.00){\circle{2.}}
\put(12.,40.){\vector(1,0){1.}}
\put(17.00,40.00){\circle{2.}}
\put(18.00,40.00){\circle{2.}}
\put(19.00,40.00){\circle{2.}}
\put(21.,40.){\vector(1,0){1.}}
\put(25.,40.){\makebox(0,0)[cc]{$\s \ldots$}}
\put(30.00,40.00){\circle{2.}}
\put(31.00,40.00){\circle{2.}}
\put(32.00,40.00){\circle{2.}}
\put(33.00,40.00){\circle{2.}}
\put(34.00,40.00){\circle{2.}}
\put(39.00,40.00){\circle{2.}}
\put(40.00,40.00){\circle{2.}}
\put(41.00,40.00){\circle{2.}}
\put(42.00,40.00){\circle{2.}}
\put(43.00,40.00){\circle{2.}}
\put(44.00,40.00){\circle{2.}}
\put(45.00,40.00){\circle{2.}}
\put(45.00,41.00){\circle{2.}}
\put(45.00,42.00){\circle{2.}}
\put(45.00,43.00){\circle{2.}}
\put(45.00,44.00){\circle{2.}}
\put(45.00,45.00){\circle{2.}}
\put(45.00,46.00){\circle{2.}}
\put(45.00,47.00){\circle{2.}}

\put(45.00,53.00){\circle{2.}}
\put(45.00,54.00){\circle{2.}}
\put(45.00,55.00){\circle{2.}}
\put(45.00,56.00){\circle{2.}}
\put(45.00,57.00){\circle{2.}}
\put(45.00,58.00){\circle{2.}}
\put(45.00,59.00){\circle{2.}}
\put(45.00,60.00){\circle{2.}}
\put(45.00,61.00){\circle{2.}}
\put(45.00,62.00){\circle{2.}}
\put(45.00,63.00){\circle{2.}}
\put(45.00,64.00){\circle{2.}}
\put(45.00,65.00){\circle{2.}}
\put(45.00,66.00){\circle{2.}}
\put(45.00,67.00){\circle{2.}}
\put(45.00,68.00){\circle{2.}}
\put(45.00,69.00){\circle{2.}}
\put(45.00,70.00){\circle{2.}}
\put(45.00,71.00){\circle{2.}}
\put(45.00,72.00){\circle{2.}}
\put(45.00,73.00){\circle{2.}}
\put(45.00,74.00){\circle{2.}}
\put(45.00,75.00){\circle{2.}}
\put(45.00,76.00){\circle{2.}}
\put(45.00,77.00){\circle{2.}}
\put(45.00,78.00){\circle{2.}}
\put(45.00,79.00){\circle{2.}}
\put(45.00,80.00){\circle{2.}}
\put(45.00,80.00){\circle{2.}}
\put(46.00,80.00){\circle{2.}}
\put(47.00,80.00){\circle{2.}}
\put(48.00,80.00){\circle{2.}}
\put(49.00,80.00){\circle{2.}}
\put(50.00,80.00){\circle{2.}}
\put(51.00,80.00){\circle{2.}}
\put(52.00,80.00){\circle{2.}}
\put(53.00,80.00){\circle{2.}}
\put(54.00,80.00){\circle{2.}}
\put(55.00,80.00){\circle{2.}}
\put(56.00,80.00){\circle{2.}}
\put(57.00,80.00){\circle{2.}}

\put(25.,20.){\makebox(0,0)[cc]{$\s \ldots$}}
\put(33.00,20.00){\circle{2.}}
\put(34.00,20.00){\circle{2.}}
\put(35.00,20.00){\circle{2.}}
\put(36.00,20.00){\circle{2.}}
\put(37.00,20.00){\circle{2.}}
\put(38.00,20.00){\circle{2.}}
\put(39.00,20.00){\circle{2.}}
\put(40.00,20.00){\circle{2.}}
\put(41.00,20.00){\circle{2.}}
\put(42.00,20.00){\circle{2.}}
\put(43.00,20.00){\circle{2.}}
\put(44.00,20.00){\circle{2.}}
\put(45.00,20.00){\circle{2.}}
\put(46.00,20.00){\circle{2.}}
\put(47.00,20.00){\circle{2.}}
\put(48.00,20.00){\circle{2.}}
\put(49.00,20.00){\circle{2.}}
\put(50.00,20.00){\circle{2.}}
\put(51.00,20.00){\circle{2.}}
\put(52.00,20.00){\circle{2.}}
\put(53.00,20.00){\circle{2.}}
\put(54.00,20.00){\circle{2.}}
\put(55.00,20.00){\circle{2.}}
\put(56.00,20.00){\circle{2.}}
\put(57.00,20.00){\circle{2.}}
\put(58.00,20.00){\circle{2.}}
\put(59.00,20.00){\circle{2.}}
\put(60.00,20.00){\circle{2.}}
\put(61.00,20.00){\circle{2.}}
\put(112.00,70.00){\circle{2.}}
\put(113.00,70.00){\circle{2.}}
\put(117.00,70.00){\circle{2.}}
\put(118.00,70.00){\circle{2.}}
\put(119.00,70.00){\circle{2.}}
\put(120.00,70.00){\circle{2.}}
\put(122.,70.){\vector(1,0){1.}}
\put(129.00,70.00){\circle{2.}}
\put(130.00,70.00){\circle{2.}}
\put(127.00,70.00){\circle{2.}}
\put(128.00,70.00){\circle{2.}}
\put(132.,70.){\vector(1,0){1.}}
\put(136.,70.){\makebox(0,0)[cc]{$\s \ldots$}}
\put(142.00,70.00){\circle{2.}}
\put(143.00,70.00){\circle{2.}}
\put(144.00,70.00){\circle{2.}}
\put(148.00,70.00){\circle{2.}}
\put(149.00,70.00){\circle{2.}}
\put(150.00,70.00){\circle{2.}}
\put(151.00,70.00){\circle{2.}}
\put(152.00,70.00){\circle{2.}}
\put(153.00,70.00){\circle{2.}}
\put(154.00,70.00){\circle{2.}}
\put(155.00,70.00){\circle{2.}}
\put(156.00,70.00){\circle{2.}}
\put(157.00,70.00){\circle{2.}}
\put(158.00,70.00){\circle{2.}}
\put(159.00,70.00){\circle{2.}}
\put(160.00,70.00){\circle{2.}}
\put(161.00,70.00){\circle{2.}}
\put(162.00,70.00){\circle{2.}}
\put(163.00,70.00){\circle{2.}}
\put(164.00,70.00){\circle{2.}}
\put(166.,70.){\vector(1,0){1.}}
\end{picture}
\]
The occurance of the constant matrices ${}_{vv}R_+$ will greatly simplify
the calculations of the algebraic Bethe ansatz.

In order to perform the nested algebraic Bethe ansatz (NABA) we define an
auxiliary transfer matrix
as the (super) Markov trace of the monodromy matrix, that is,
\begin{equation}
{}_{vw}\tau^{ \{ j \} }_{ \{i\} }(y)=\sum_i {}_vK^\alpha_\alpha
~~{}_{vw}U^{\alpha \{ j \} }_{\alpha \{ i \} }(y)
=
\ba{c}
\unitlength=0.50mm
\begin{picture}(95.,49.)
\put(45.,15.){\makebox(0,0)[cc]{$\cdots$}}
\put(21.,-3.){\makebox(0,0)[cc]{$\s i_1$}}
\put(31.,-3.){\makebox(0,0)[cc]{$\s i_2$}}
\put(71.,-3.){\makebox(0,0)[cc]{$\s i_k$}}
\put(21.,47.){\makebox(0,0)[cc]{$\s j_1$}}
\put(31.,47.){\makebox(0,0)[cc]{$\s j_2$}}
\put(71.,47.){\makebox(0,0)[cc]{$\s j_k$}}
\put(70.,0.){\vector(0,1){45.}}
\put(30.,0.){\vector(0,1){45.}}
\put(20.,0.){\vector(0,1){45.}}
\put(80.,35.){\vector(1,0){1.}}

\put(57.00,35.00){\circle{2.}}
\put(58.00,35.00){\circle{2.}}
\put(59.00,35.00){\circle{2.}}
\put(60.00,35.00){\circle{2.}}
\put(61.00,35.00){\circle{2.}}
\put(62.00,35.00){\circle{2.}}
\put(63.00,35.00){\circle{2.}}
\put(64.00,35.00){\circle{2.}}
\put(65.00,35.00){\circle{2.}}
\put(66.00,35.00){\circle{2.}}
\put(67.00,35.00){\circle{2.}}
\put(68.00,35.00){\circle{2.}}
\put(72.00,35.00){\circle{2.}}
\put(73.00,35.00){\circle{2.}}
\put(74.00,35.00){\circle{2.}}
\put(75.00,35.00){\circle{2.}}
\put(76.00,35.00){\circle{2.}}
\put(77.00,35.00){\circle{2.}}
\put(78.00,35.00){\circle{2.}}
\put(57.00,15.00){\circle{2.}}
\put(58.00,15.00){\circle{2.}}
\put(59.00,15.00){\circle{2.}}
\put(60.00,15.00){\circle{2.}}
\put(61.00,15.00){\circle{2.}}
\put(62.00,15.00){\circle{2.}}
\put(63.00,15.00){\circle{2.}}
\put(64.00,15.00){\circle{2.}}
\put(65.00,15.00){\circle{2.}}
\put(66.00,15.00){\circle{2.}}
\put(67.00,15.00){\circle{2.}}
\put(68.00,15.00){\circle{2.}}
\put(69.00,15.00){\circle{2.}}
\put(70.00,15.00){\circle{2.}}
\put(71.00,15.00){\circle{2.}}
\put(72.00,15.00){\circle{2.}}
\put(73.00,15.00){\circle{2.}}
\put(74.00,15.00){\circle{2.}}
\put(75.00,15.00){\circle{2.}}
\put(76.00,15.00){\circle{2.}}
\put(77.00,15.00){\circle{2.}}
\put(78.00,15.00){\circle{2.}}
\put(79.00,15.00){\circle{2.}}
\put(80.00,15.00){\circle{2.}}
\put(45.,35.){\makebox(0,0)[cc]{$\cdots$}}
\put(15.00,15.00){\circle{2.}}
\put(16.00,15.00){\circle{2.}}
\put(17.00,15.00){\circle{2.}}
\put(18.00,15.00){\circle{2.}}
\put(19.00,15.00){\circle{2.}}
\put(20.00,15.00){\circle{2.}}
\put(21.00,15.00){\circle{2.}}
\put(22.00,15.00){\circle{2.}}
\put(23.00,15.00){\circle{2.}}
\put(24.00,15.00){\circle{2.}}
\put(25.00,15.00){\circle{2.}}
\put(26.00,15.00){\circle{2.}}
\put(27.00,15.00){\circle{2.}}
\put(28.00,15.00){\circle{2.}}
\put(29.00,15.00){\circle{2.}}
\put(30.00,15.00){\circle{2.}}
\put(31.00,15.00){\circle{2.}}
\put(32.00,15.00){\circle{2.}}
\put(33.00,15.00){\circle{2.}}
\put(34.00,15.00){\circle{2.}}
\put(35.00,15.00){\circle{2.}}
\put(36.00,15.00){\circle{2.}}
\put(37.00,15.00){\circle{2.}}
\put(15.00,35.00){\circle{2.}}
\put(16.00,35.00){\circle{2.}}
\put(17.00,35.00){\circle{2.}}
\put(18.00,35.00){\circle{2.}}
\put(22.00,35.00){\circle{2.}}
\put(23.00,35.00){\circle{2.}}
\put(24.00,35.00){\circle{2.}}
\put(25.00,35.00){\circle{2.}}
\put(26.00,35.00){\circle{2.}}
\put(28.00,35.){\vector(1,0){1.}}

\put(32.00,35.00){\circle{2.}}
\put(33.00,35.00){\circle{2.}}
\put(34.00,35.00){\circle{2.}}
\put(35.00,35.00){\circle{2.}}
\put(36.00,35.00){\circle{2.}}
\put(37.00,35.00){\circle{2.}}
\put(39.00,35.){\vector(1,0){1.}}

\put(5.00,25.){\circle*{2.5}}

\put(5.00,25.00){\circle{2.}}
\put(6.00,26.00){\circle{2.}}
\put(7.00,27.00){\circle{2.}}
\put(8.00,28.00){\circle{2.}}
\put(9.00,29.00){\circle{2.}}
\put(10.00,30.00){\circle{2.}}
\put(11.00,31.00){\circle{2.}}
\put(12.00,32.00){\circle{2.}}
\put(13.00,33.00){\circle{2.}}
\put(14.00,34.00){\circle{2.}}
\put(15.00,35.00){\circle{2.}}

\put(5.00,25.00){\circle{2.}}
\put(6.00,24.00){\circle{2.}}
\put(7.00,23.00){\circle{2.}}
\put(8.00,22.00){\circle{2.}}
\put(9.00,21.00){\circle{2.}}
\put(10.00,20.00){\circle{2.}}
\put(11.00,19.00){\circle{2.}}
\put(12.00,18.00){\circle{2.}}
\put(13.00,17.00){\circle{2.}}
\put(14.00,16.00){\circle{2.}}
\put(15.00,15.00){\circle{2.}}

\put(85.00,10.00){\circle{2.}}
\put(84.00,11.00){\circle{2.}}
\put(83.00,12.00){\circle{2.}}
\put(82.00,13.00){\circle{2.}}
\put(81.00,14.00){\circle{2.}}
\put(80.00,15.00){\circle{2.}}

\put(85.00,40.00){\circle{2.}}
\put(84.00,39.00){\circle{2.}}
\put(83.00,38.00){\circle{2.}}
\put(82.00,37.00){\circle{2.}}
\put(81.00,36.00){\circle{2.}}
\put(80.00,35.00){\circle{2.}}

\put(85.00,40.00){\circle{2.}}
\put(86.00,39.00){\circle{2.}}
\put(87.00,38.00){\circle{2.}}
\put(88.00,37.00){\circle{2.}}
\put(89.00,36.00){\circle{2.}}
\put(90.00,35.00){\circle{2.}}
\put(91.00,34.00){\circle{2.}}
\put(92.00,33.00){\circle{2.}}
\put(93.00,32.00){\circle{2.}}
\put(94.00,31.00){\circle{2.}}
\put(95.00,30.00){\circle{2.}}
\put(96.00,29.00){\circle{2.}}
\put(97.00,28.00){\circle{2.}}
\put(98.00,27.00){\circle{2.}}
\put(99.00,26.00){\circle{2.}}
\put(100.00,25.00){\circle{2.}}

\put(85.00,10.00){\circle{2.}}
\put(86.00,11.00){\circle{2.}}
\put(87.00,12.00){\circle{2.}}
\put(88.00,13.00){\circle{2.}}
\put(89.00,14.00){\circle{2.}}
\put(90.00,15.00){\circle{2.}}
\put(91.00,16.00){\circle{2.}}
\put(92.00,17.00){\circle{2.}}
\put(93.00,18.00){\circle{2.}}
\put(94.00,19.00){\circle{2.}}
\put(95.00,20.00){\circle{2.}}
\put(96.00,21.00){\circle{2.}}
\put(97.00,22.00){\circle{2.}}
\put(98.00,23.00){\circle{2.}}
\put(99.00,24.00){\circle{2.}}
\put(100.00,25.00){\circle{2.}}
\end{picture}
\ea\ ,
\end{equation}

where
\begin{eqnarray}
{}_vK&=& \left( \begin{array}{clcr}
1& 0 & 0\\
0 & q^2 & 0 \\
0 & 0 & -q^2
\end{array}\right).\label{kmatrix}
\end{eqnarray}
Therefore the ${}_{vw}\tau(y)$ form a one-parameter family of commuting
operators and it may be shown that they commute with the transfer matrix,
${}_{ww}\tau(y)$ of the Hamiltonian (\ref{ham1}).  This means that they have
a common set of
eigenvectors.  We find that
\begin{eqnarray}
{}_{vw}\tau
(y)=~~{}_{vw}U_1^1(y)+q^2~~{}_{vw}U_2^2(y)-q^2~~{}_{vw}U_3^3(y).\label{vwt}
\end{eqnarray}

Take the lowest weight state as a reference state (pseudo-vacuum) in $W$,
which we denote as $|0>_i$.  
Then $|0>=\otimes_{i=1}^k |0>_i$ and we find the action of
the doubled monodromy matrix on this reference state to be given by
\begin{eqnarray}
{}_{vw}U(x)_k|0>=
\left(\begin{array}{clcr}
I(x)^k  & 0 & 0 \\
0 & I(x)^k & 0 \\
{}_{vw }U^3_1(x) & {}_{vw }U^3_2(x) & 1
\end{array}\right)|0>, \label{vwaction}
\end{eqnarray}
where
$$I(x)= \frac {(1-x q^{\alpha})}{(1-x q^{-\alpha-2} )}.$$

We construct a set of eigenstates of the transfer matrix using
the technique of the NABA.
The creation operators are ${}_{vw}U^3_1(x)$, ${}_{vw}U^3_2(x)$
due to the choice of reference state.  Thus we use the following
for the ansatz for the eigenstates of ${}_{vw}\tau(y)$:
\begin{eqnarray}
\Psi={}_{vw}U^3_{a_1}(x_1)~~{}_{vw}U^3_{a_2}(x_2)...{}_{vw}U^3_{a_r}(x_r)~~
\Psi^{(1)}_{ \{a\} }|0> ,\label{ansatz}
\end{eqnarray}
where indices $a_i$ have values 1 or 2.  We seek a solution of the
eigenvalue equation
\begin{eqnarray}
{}_{vw}\tau(y)\Psi = {}_{vw}\Lambda(y) \Psi.\label{eig}
\end{eqnarray}
The action of these states is determined by the monodromy matrix and the
relations (\ref{newybe}). The relations
necessary for the construction of the NABA are
\begin{eqnarray}
{}_{vw}U^3_3(y)~~{}_{vw}U^3_\alpha(x)&=&-\frac
{1}{q~E(y/x)}~~{}_{vw}U^3_\alpha(x)~~{}_{vw}U^3_3(y) -
\frac{y~C( y/x)}{x~q~E( y/x)}~~{}_{vw}U^3_\alpha(y)~~{}_{vw}U^3_3(x)\nonumber \\
&& -\left(\frac{q-q^{-1}}{q}\right)\sum_\beta {}_{vw}U^3_\alpha(y)
~~{}_{vw}U^\beta_\alpha(x),\\
{}_{vw}U^\gamma_\beta(y)~~ {}_{vw}U^3_\alpha(x)&=&\frac
{r_+{}^{\gamma\alpha' }_{\delta'\gamma'}~~
r^{\beta'\gamma'}_{\beta \alpha} (x/y)} {q~E(x/y)} ~~{}_{vw}U^3_{\alpha'} (x)
~~{}_{vw}U^{\delta'}_{\beta'}(y) 
- \frac{ x~~r_+{}^{\gamma\alpha'}_{\delta'\beta}~~ C(x/y) }{y~ q~ E(x/y)}
~~{}_{vw}U^3_{\alpha'} (y)
~~{}_{vw}U^{\delta'}_{\alpha}(x), \label{vwrels}
\end{eqnarray}
with the indices taking values of 1 and 2.
It can be seen that this $R$-matrix $r(y)$ also fulfills a Yang-Baxter
equation and can be identified with  the
$R$-matrix of the quantum spin $\frac 12$ Heisenberg (XXZ) model.
The action of the ansatz (\ref{ansatz}) on the diagonal elements of the
monodromy matrix (\ref{mono}) is given by
\begin{eqnarray}
{}_{vw}U^3_3(y)\Psi &=&\frac{(-1)^r}{q^r}\prod^r_{i=1}\frac 1{E(y/x_i)}\Psi
+ \mbox{ u.t.}\nonumber
\end{eqnarray}
\begin{eqnarray}
[{}_{vw}U^1_1(y)+q^2~~{}_{vw}U^2_2(y)]\Psi &=&
\frac{I(y)^k}{q^r}\prod_{j=1}^r \frac 1{E(x_j/y)}
\prod_{l=1}^r {}_{vw}U^3_{b_l}(x_l)|0> q \tau^{(1)}(y)^{b_1...b_r}\Psi^{(1)} +\mbox{ u.t.}\nonumber 
\end{eqnarray}
where
\begin{eqnarray}
{}_{vw}\tau^{(1)}(y) =
q^{-1}~~{}_{vw}U^{(1)}{}_1^1(y)+q~~{}_{vw}U^{(1)}{}_2^2(y).\label{vwt1}
\end{eqnarray}

In order that the eigenvalue problem (\ref{eig}) is satisfied, it is
necessary to solve a new eigenvalue
problem for the nesting as follows:
$${}_{vw}\tau^{(1)}(y)\Psi^{(1)}=\Lambda^{(1)}(y,\{y_j\})\Psi^{(1)},$$
where
\begin{eqnarray}
\Psi^{(1)}={}_{vw}U^{(1)}{}^2_1(y_1)~~{}_{vw}U^{(1)}{}^2_1(y_2)
...{}_{vw}U^{(1)}{}^2_1(y_m)|0>^{(1)}.\label{nansatz}
\end{eqnarray}
The second level reference state is given by $|0>^{(1)}=\otimes_{i=1}^r
|2>_i$.
We represent the nested monodromy matrix as
\begin{eqnarray}
{}_{vw}U^{(1)}_m(y) =\left(\begin{array}{clcr}
{}_{vw}U^{(1)}{}_1^1(y) & {}_{vw}U^{(1)}{}_2^1(y)\\
{}_{vw}U^{(1)}{}^2_1(y) & {}_{vw}U^{(1)}{}_2^2(y)
\end{array}\right).
\end{eqnarray}
The action of the nested monodromy matrix ${}_{vw}U^{(1)}_m(y)$ on the
second level reference state is
\begin{eqnarray}
{}_{vw}U^{(1)}{}_1^1(y)|0>^{(1)}&=&q^r\prod^r_{j=1}E(x_j/y)|0>^{(1)},
\nonumber\\
{}_{vw}U^{(1)}{}_2^2(y)|0>^{(1)}&=&\prod^r_{j=1} A(x_j/y)|0>^{(1)}.\nonumber
\\
\end{eqnarray}

The action of ${}_{vw}\tau^{(1)}(y)$ on the ansatz (\ref{nansatz}) is
computed similar to the
first level case from the relations (\ref{newybe}).  We obtain
\begin{eqnarray}
{}_{vw}U^{(1)}{}_1^1(y)\Psi^{(1)}&=&q^{r-m}\prod^m_{i=1}
\frac{A(y_i/y)}{E(y_i/y)}\prod_{l=1}^r E(x_l/y) \Psi^{(1)}+
\mbox{ u.t. }\nonumber \\
{}_{vw}U^{(1)}{}_2^2(y)\Psi^{(1)}&=&q^m\prod^m_{i=1}
\frac{A(y/y_i)}{E(y/y_i)}\prod_{l=1}^r A(x_l/y) \Psi^{(1)}+
\mbox{ u.t. }\nonumber \\
\end{eqnarray}

The eigenvalues for the auxilliary transfer matrix, ${}_{vw}\tau(y)$ are
found to be
\begin{eqnarray}
{}_{vw}\Lambda(y)&=&q^{-m} I(y)^k \prod^m_{i=1}
\frac{ A(y_i/y)}{E(y_i/y)}+q^{2+m-r} I(y)^k \prod_{i=1}^r
\frac{ A(x_i/y)}{E(x_i/y)} \prod^m_{j=1}\frac{A(y/y_j)}{E(y/y_j)}
-(-1)^r q^{2-r} \prod^r_{i=1} \frac 1{E(y/x_i)},\label{eigs}
\end{eqnarray}
provided the ``unwanted terms" cancel.  The cancellation of these terms lead
to the Bethe ansatz equations obtained by eliminating the poles of
the eigenvalues (\ref{eigs})
\begin{eqnarray}
\prod_{i\neq n}^m \frac {qy_i-q^{-1}y_n}{q^{-1}y_i-qy_n}
&=& q^{2(1+m)-r}\prod_{j=1}^r \frac{q^{-1}y_n -qx_j}{y_n -x_j}
~~~~n=1,...,m,\nonumber \\
q^mI(x_l)^k& =&\prod^m_{j=1} \frac{y_j-x_l}{q^{-1}y_j -x_lq}
~~~~l=1,...,r.
\label{BAE5}
\end{eqnarray}
Associated with these solutions, the energies 
of the Hamiltonian are given by
$$ E= \sum_j \frac{-(q^{\alpha+1} - q^{-\alpha-1})^2}{(q^{\alpha/2}x_j^{-1/2} -
q^{-\alpha/2} x_j^{1/2})
( q^{-\alpha/2-1} x_j^{-1/2}  - q^{\alpha/2+1} x_j^{1/2})} + k \left(
{q^{\alpha+1} + q^{-\alpha-1}  }\right).$$
This expression reduces to the normal periodic case \cite{periodic} in the
rational limit as $q\rightarrow 1$.

\section{Conclusion}

In this work, we have constructed a quantum algebra invariant supersymmetric
$U$ model on a closed lattice and derived the Bethe ansatz equations.
Notice that in the Bethe ansatz equations (\ref{BAE5}) the presence of $``q"$ terms in
comparison with the corresponding equations for the usual periodic boundary
conditions \cite{meqaba}.  In fact, this feature also appeared in other
models \cite{GPPR,karow,Angi,lima} and seems to be a peculiarity of
quantum-group-invariant closed spin chains. In the limit for $q\rightarrow
1$,
the usual Bethe ansatz equations for the periodic chain \cite{meaba} are recovered.

An appealing direction for further study of the present closed
supersymmetric $U$ model with $U_q[gl(2|1)]$  invariance would be to
investigate its thermodynamic properties.  In particular the partition
function in the finite size scaling limit which can
be used to derive the operator content \cite{GPPR} of the related
statistical models.

\section*{Acknowledgements}
Katrina Hibberd, Itzhak Roditi and Angela Foerster are financially
supported by CNPq (Conselho Nacional de Deservolvimento Cient\'{\i}fico
e Tecnol\'ogico).  Itzhak Roditi also wishes to thank PRONEX/FINEP/MCT.
Jon Links is supported by the Australian Research Council.


\begin{thebibliography}{99}

\bibitem{korbook}  V.E. Korepin, F.H.L. Essler, eds., {\it ``Exactly
solvable models of stongly correlated electrons,"} (World Scientific, 1994).

\bibitem{EKS} F.H.L. Essler, V.E. Korepin, K. Schoutens, {\it Phys. Rev.
Lett. }{\bf 70}, 73 (1993).



\bibitem{ww} A.J. Bracken, M.D. Gould, J.R. Links, Y.-Z. Zhang,
{\it Phys. Rev. Lett.} {\bf 74}, 2768 (1995).

\bibitem{kul} P. Kulish, E. Sklyanin, {\it Lect. Notes Phys.} {\bf 151}, 61
(1982).

\bibitem{periodic} G. Bedurftig, H. Frahm, {\it J. Phys.} {\bf A28},
4453
(1995).

\bibitem{meaba} K.E. Hibberd, M.D. Gould, J.R. Links, {\it Phys. Rev.}
{\bf B54}, 8430 (1996).

\bibitem{marcio} P.B. Ramos and M.J. Martins, {\it Nucl.
Phys.} {\bf B474}, 678 (1996).
\bibitem{frahm} M. Pfannm\"uller and H. Frahm, {\it Nucl. Phys.} {\bf B479}, 
574 (1996); \\
M. Pfannmuller and H. Frahm, {\it J. Phys.} {\bf A30}, L543 (1997).

\bibitem{bar} R.Z. Bariev, A. Kl{\" u}mper, J.Zittartz, {\it Europhys.
Lett.}
{\bf 32}, 85 (1995).

\bibitem{me} M.D. Gould, K.E. Hibberd, J.R. Links, Y.-Z. Zhang,
{\it Phys. Lett.} {\bf A212}, 156 (1996).

\bibitem{angobc} A. Foerster, M. Karowski, {\it Nucl. Phys.} {\bf B408}, 412
(1993).

\bibitem{mass} Z. Maassarani, {\it J. Phys. }{\bf A28}, 1305
(1995).

\bibitem{meqaba} K.E. Hibberd, M.D. Gould, J.R. Links, {\it J. Phys.} {\bf
A29}, 8053
(1996).

\bibitem{martin} P.P. Martin, {\it ``Potts models and related problems in
statistical mechanics,"}
(Singapore: World Scientific, 1991).

\bibitem{GPPR} H. Grosse, S. Pallua, P. Prester  and E. Raschhofer,
{\it J. Phys. }{\bf A27}, 4761 (1994); \\
S. Pallua and P. Prester, {\it J. Phys.} {\bf A29}, 1187 (1996); \\
S. Pallua and P. Prester, {\it Phys. Rev.} {\bf D58}, 127901 (1998).

\bibitem{karow} M. Karowski, A. Zapletal, {\it Nucl. Phys. }{\bf B419}, 567
(1994).

\bibitem{kz} M. Karowski and A. Zapletal, {\it J. Phys. }{\bf A27}, 7419
(1994).

\bibitem{Angi} A. Foerster, {\it J. Phys.} {\bf A29}, 7625 (1996).

\bibitem{Links} J. Links, A. Foerster, {\it J. Phys.} {\bf A30}, 2483
(1997).

\bibitem{Linkss} J. Links, A. Foerster, M. Karowski, {\it J. Math. Phys.}
{\bf40}, 726 (1999).

\bibitem{for} A. Foerster, M. Karowski, {\it Nucl. Phys.} {\bf B408}, 512
(1993).

\bibitem{vecrep} A.J. Bracken, M.D. Gould, R.B. Zhang, {\it Mod. Phys.
Lett. }{\bf A5}, 831 (1990).

\bibitem{22} I. Cherednik, {\it Theor. Mat. Fiz. }{\bf 61}, 35 (1984).

\bibitem{23} J.R. Links, M.D. Gould, {\it Int. J. Mod. Phys. }{\bf B10},
3461 (1996).


\bibitem{lima} A. Lima-Santos  and R.C.T. Ghiotto, {\it J. Phys.}
{\bf A31}, 505 (1998);\\
A. Lima-Santos, {\it Nucl. Phys. }{\bf B522}, 503 (1998).



\end{thebibliography}
\end{document}